\documentclass[prx,twocolumn,superscriptaddress,showpacs]{revtex4}

\usepackage{amsmath,epsfig}

\newcommand{\qq}{q^\ast}
\newcommand{\friction}{\gamma}
\newcommand{\viscosity}{\eta}

\begin{document}

\title{Particle sorting by a structured microfluidic 
ratchet device with tunable selectivity: Theory and Experiment}

\date{\today}

\author{Lukas Bogunovic}
\affiliation{Universit\"at Bielefeld, Fakult\"at f\"ur Physik, 33615 Bielefeld, Germany}
\author{Ralf Eichhorn}
\affiliation{NORDITA, Roslagstullsbacken 23, 106 91 Stockholm, Sweden}
\author{Jan Regtmeier}
\affiliation{Universit\"at Bielefeld, Fakult\"at f\"ur Physik, 33615 Bielefeld, Germany}
\author{Dario Anselmetti}
\affiliation{Universit\"at Bielefeld, Fakult\"at f\"ur Physik, 33615 Bielefeld, Germany}
\author{Peter Reimann}
\email{reimann@Physik.Uni-Bielefeld.DE}
\affiliation{Universit\"at Bielefeld, Fakult\"at f\"ur Physik, 33615 Bielefeld, Germany}

\begin{abstract}
We theoretically predict and experimentally 
demonstrate that several different particle species
can be separated from each other by means of a ratchet 
device, consisting of periodically arranged
triangular (ratchet) shaped obstacles.
We propose an explicit algorithm for
suitably tailoring the externally applied,
time-dependent voltage protocol so that 
one or several, arbitrarily selected particle species   
are forced to migrate oppositely to all the remaining 
species.
As an example we present numerical simulations for a mixture of 
five species, labelled according to their increasing 
size, so that species 2 and 4 simultaneously move in
one direction and species 1, 3, and 5 in the other.
The selection of species to be separated 
from the others can be changed at any time by
simply adapting the voltage protocol.
This general theoretical concept to utilize 
one device for many different sorting tasks 
is experimentally confirmed for a mixture 
of three colloidal particle species.
\end{abstract}



\maketitle

\section{Introduction}
Ratchet effects refer to directed transport 
under non-equilibrium conditions
in periodic systems with broken spatial symmetry,
and are currently attracting considerable attention
in a wide variety of different fields \cite{871,361,125,872}.
So far, the majority of pertinent works 
has been devoted to basic theoretical concepts, 
fueled -- among others -- †by the exciting perspective 
to exploit ratchet effects for particle sorting 
purposes at the nano- and micro-scale.
Concerning experimental ratchet devices, 
directed transport {\em per se} has been 
demonstrated by now in quite a number of systems
as well, see refs. 
\cite{452,453,873,280,874,888,157,875,876,877,570,878,890}
for just a few examples.
On the other hand, the actual sorting of different 
particle species has been achieved only in a 
few experimental systems \cite{879,880,405,881,136,177,35,cho05,882,889,883,884}. 
All of them have been conducted in some
microstructures with broken spatial symmetry, 
while the indispensable non-equilibrium 
conditions have been implemented mostly
by utilizing electrokinetic 
effects \cite{879,880,405,881,136,177,35,cho05,882,889}, 
but also via magnetic fields \cite{883}
or intracellular morphological 
changes \cite{884}.

Along these lines of 
experimentally realizing ratchet effects by exploiting electrokinetics
\cite{pet97,gas02},
our present work addresses the natural next question:
Given a mixture of different particle species, 
is it possible to separate any arbitrarily selected subset 
of species from all the others with the 
help of one and the same microstructured 
ratchet device?
After setting the general
framework we have in mind, we will
outline the basic theoretical ideas,
verify them quantitatively by numerical simulations,
and finally provide the experimental 
proof-of-principle for a colloidal
suspension containing three different species
of polystyrene beads.

The general experimental framework is 
schematically illustrated in Fig.~1
(see also Appendix A.1).
The central, topographically structured area 
consists of periodically arranged, triangular 
(ratchet) shaped obstacles.
The entire device is filled with a buffer 
solution, containing particles in sufficiently
high dilution, so that their mutual 
interaction is negligible.
It is actuated by an external voltage
like indicated in Fig.~1a.
\begin{figure}
\epsfxsize=\columnwidth
\epsfbox{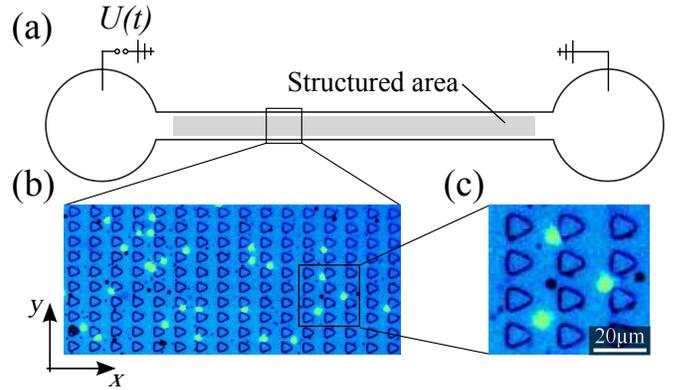} 
\caption{Experimental setup.
(a): Schematic top view ($x$-$y$-plane) of the 
entire device (not to scale) with 
reservoirs and electrodes at its ends 
and a central, microstructured part (grey).
(b) Optical micrograph image of the microstructured
area. The entire microstructured region extends 
over $10000\,\mu$m$\times 300\,\mu$m 
in the $x$-$y$-plane and 10$\,\mu$m in height 
($z$-direction).
(c) Magnification, exhibiting periodically arranged
obstacles (posts) of ratchet-shaped cross section
(see Fig.~2 for the dimensions of the obstacles)
with a period of $20\,\mu\mbox{m}$ in $x$ direction
and $14\,\mu\mbox{m}$ in $y$ direction.
Also visible are microbeads of three different diameters
(2.9$\, \mu$m, 1.9$\, \mu$m, and 1.1$\, \mu$m).
The largest and smallest beads can be easily distinguished 
by their size.
The medium sized beads are fluorescently labelled and thus
appear as bright spots.}
\label{fig1}
\end{figure}

\section{Theoretical Concepts}
%
%
\subsection{Electrokinetic effects}
Upon applying an electrical DC- and AC-voltage $U(t)$ 
of the general form
\begin{equation}
U(t)=U_\mathrm{DC}+U_\mathrm{AC}\,\sin(\omega t) 
\label{1}
\end{equation}
to the electrodes in Fig.~1a,
an electrical field $\vec E(\vec r,t)$ is created
at position $\vec r=(x,y,z)$ and time $t$ within the
device.
For all experimentally realistic DC- and AC-voltages
in (\ref{1}), the electrostatic approximation is valid \cite{reg11},
so that the field induced by $U(t)$ is
\begin{equation}
\vec E(\vec r,t)=U(t)\, \vec E_1(\vec r)/U_1 \ ,
\label{6}
\end{equation}
where $\vec E_1(\vec r)$ denotes the electrical field
induced by a unit DC-voltage $U_1:=1\,$V. 
To determine $\vec E_1(\vec r)$
within the structured region (Fig.\ 1b), we numerically 
solved Laplace's equation with Neumann boundary conditions,
thus approximately treating the microstructure 
as a perfect isolator \cite{reg11,834}. 
The resulting electrical field $\vec E_1(\vec r)$
around an experimentally realistic obstacle is illustrated
in Fig.\ 2a.
Outside the structured region, an approximative
treatment along the lines of \cite{433} 
was adopted to estimate the voltage drop
over the structured region, given the voltage 
$U_1:=1\,$V is applied to the device electrodes
(cf.\ Fig.\ 1).
\begin{figure}
\epsfxsize=\columnwidth
\epsfbox{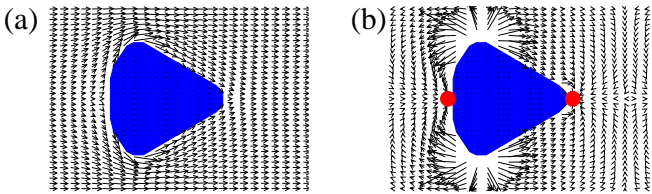} 
\caption{Force fields.
(a) Electrical field $\vec E_1(\vec r)$
around one idealized obstacle
within a region of size $20\,\mu\mbox{m} \times 14\,\mu\mbox{m}$,
corresponding to one basic ``unit cell'' of
the periodic microstructure,
see also Fig.\ 1c and Eq.\ (\ref{6}).
The shape of the obstacle is chosen to
closely approximate the real experimental shape of the
triangular posts as formed by the soft
lithography process
(see Appendix A.1) and
as characterized by REM imaging.
The obstacle extends $8.5\,\mu\mbox{m}$ from
baseline to tip ($x$ direction) and
$8.5\,\mu\mbox{m}$ from side vertex to side vertex
($y$ direction).
(b) Dielectrophoretic force field 
$\alpha\, \vec\nabla\vec E_1^2(\vec r)/2$, 
under the assumption that $\alpha <0$, 
see also Eq.\ (\ref{3}) and Appendix A.5.
Red dots indicate particle traps 
(minima of the potential (\ref{3})).
Both fields inherit the spatial periodicity
of the microstructure (Fig.~1b) and are
depicted here in arbitrary units.
}
\label{fig2}
\end{figure}

Due to the electric field $\vec E(\vec r,t)$,
a particle in the microstructure experiences
two different electrokinetic effects \cite{gas02,squ05}:
First, it is driven by electrophoresis.
The particle velocity $\dot{\vec r}$ due to purely
electrophoretic forces (i.e. neglecting
other electrokinetic effects, thermal noise etc.)
can be expressed in terms of
the fluid's permittivity $\epsilon$,
its viscosity $\viscosity$, and the particle's zeta potential 
$\zeta$ by the well-known 
Helmholtz-Smoluchowski relation 
\begin{equation}
\dot{\vec r} = \vec E(\vec r,t) \epsilon\zeta/\viscosity \, ,
\label{7}
\end{equation}
provided the Debye screening length of
the counter-ions is much smaller than the
particle diameter \cite{squ05,mor70}.
The latter condition can be safely taken 
for granted in view of a Debye length 
of $\approx 10\,$nm for our experimental
10 mM phosphate buffer \cite{squ05}.
The Helmholtz-Smoluchowski relation (\ref{7}) applies
to particles of any shape \cite{mor70}
and holds also if the external electric field
varies over distances comparable to the
particle size \cite{keh85}.
We further note that it can formally
be rewritten as a force balance
$\friction\, \dot{\vec r}=\qq \vec E(\vec r)$
between the viscous (Stokes) friction force
$\friction\, \dot{\vec r}$ with friction coefficient
$\gamma$ and 
an effective electrostatic force $\qq \vec E(\vec r)$
with an ``\emph{effective} charge'' $\qq$
defined as
\begin{equation}
\qq := \friction\, \epsilon \zeta /\viscosity
\label{2}
\end{equation}
This quantity $\qq$
is a purely formal abbreviation
which has the dimension of a charge but 
has very little to do with any kind of 
actual physical charge.

Second, the electrical field $\vec E(\vec r,t)$
induces a dipole moment $\vec p$ in the particle
and the ion cloud around it,
which is characterized by the relevant (effective)
polarizability $\alpha$.
In general, $\alpha$ is a complex, frequency dependent
quantity, and the dipole moment $\vec p$ may be composed
of several contributions according to the frequency components
contained in the electrical field $\vec E(\vec r, t)$.
Under our present experimental conditions, however,
the quasi-static approximation with $\alpha$ referring
to the zero frequency limit can be invoked,
since polarization
processes are much faster than time-variations of
the electric field \cite{749}.
Likewise, conductive and dielectric losses can be
safely neglected, so that $\alpha$ is given as
a real-valued quantity.
The dipole moment $\vec p$ is thus of the
form $\alpha \vec E(\vec r,t)$, and
the standard dipole-field-coupling force
$(\vec p\cdot \vec\nabla)\vec E$ derives 
from a dielectrophoretic potential
\begin{equation}
W(\vec r,t) = -\alpha \vec E^2(\vec r,t)/2 \ .
\label{3}
\end{equation}
Figure 2b depicts the corresponding
dielectrophoretic force field
around an experimentally realistic obstacle.
Similar as $\qq$, also the polarizability
$\alpha$ is an effective quantity depending on
particle and solution properties. A common
approximation for the present situation of a
colloidal bead (diameter $d$) suspended in an ionic buffer solution is
\begin{equation}
\alpha = -2\pi\epsilon(d/2)^3 \, ,
\label{8}
\end{equation}
obtained from the well-known basic formula
for the polarization of a non-conducting dielectric sphere 
in a conducting dielectric medium with permittivity $\epsilon$
at low frequencies 
of the electric field \cite{749}.
\subsection{Particle dynamics}
In addition to the two electrokinetic effects
from the previous Section,
other relevant forces acting on the particle are
thermal fluctuations and
interactions with the microstructure,
while inertia effects are so tiny
that they do not play any noticeable role \cite{squ05,purcell77}.
We model the particle dynamics by the standard overdamped
Langevin equation
\cite{871,361,125,872,879,880,405,881,136,177,35,882,889,834,reg11}
\begin{equation}
\friction\,\dot{\vec r} = 
\qq \vec E(\vec r,t) + \alpha \vec{\nabla}\vec E^2(\vec r,t)/2+
\vec F(\vec r)+ \vec{\xi} (t) \ .
\label{5}
\end{equation} 
On the right hand side, the first and second terms are 
the electrophoretic and dielectrophoretic forces
as detailed above.
The third term $\vec F(\vec r)$ accounts for the repulsion by 
the microstructure, idealized as hard wall forces due to
the periodic obstacles from Fig.\ 2 and the channel walls.
The last term $\vec\xi (t)$ describes thermal fluctuations.
Its components $\xi_\sigma(t)$, $\sigma\in\{x,y,z\}$, are modeled as 
usual \cite{871,361,125,872}
by independent Gaussian noises of zero average and
correlation $\langle \xi_\sigma(t)\,\xi_\sigma(s)\rangle = 2\friction k T \delta(t-s)$,
where $k$ is Boltzmann's constant, $T\approx 290\,$K the ambient temperature,
and where $\langle\cdot\rangle$ denotes the average over
many realizations of the Gaussian noise.
The quantitative values of the remaining model parameters 
$\friction$, $\qq$, and $\alpha$ 
are provided in the Appendices A.2-A.4.

The model (\ref{5}) follows the standard approach
for describing electrokinetic effects and thus
neglects non-linear terms
of cubic and higher order in the external electrical field
$\vec E(\vec r,t)$,
which would, in particular, take into account local
changes of $\vec E(\vec r,t)$ due to the
presence of the particle.
Likewise, hydrodynamic interactions are neglected.
This approximation is justified in the present case
of electrophoretically driven particle motion, because
in electrophoresis hydrodynamic interactions are 
essentially screened \cite{long96}. Remaining hydrodynamic
effects, in particular from interactions of the particle
with the channel structure, are approximately taken into
account by determining the friction coefficient $\gamma$
from the particle diffusion
in the microstructure, see Appendix A.2.
Since the friction forces
are dominated by hydrodynamic interactions with
the bottom and top
walls of the microfluidic channel
(which are isotropic
in $x$ and $y$ direction),
a possible orientation dependence of the friction
force in the $x$-$y$ plane for particle
movements parallel or perpendicular to the obstacle walls
is expected to be negligibly small, and is thus
omitted in our model.
Ultimately, all these approximation are justified by
the good agreement between the theoretical predictions
and the experimental results (see, e.g., Figs.~5a and 6). 
\subsection{Separation principle}
As already mentioned we want to address the problem
of separating an arbitrarily selected subset of particle species
from a mixture of many different particle species.
Obviously, the ``optimal'' realization of such a
separation is to move the selected
subset into one direction along the microstructured channel,
while, at the same time, all the remaining species migrate into the opposite
direction. In other words, the average particle velocity
in $x$ direction
\begin{equation}
v := \lim_{t \to \infty} \frac{\langle x(t)-x(0) \rangle}{t}
\label{eq:v}
\end{equation}
assumes different signs for the two sets of particle species.
In the following, we devise the theoretical idea how
this task can be accomplished with our microfluidic
device from Fig.~1 by applying a periodic sequence
of combined DC- and AC-voltage pulses.

To keep things as simple as possible, we focus on
cases where the DC-voltage mainly determines the
electrophoresis, while the AC-voltage predominantly
rules the dielectrophoresis.
More precisely, we restrict ourselves to DC- and AC-voltages in (\ref{1})
such that practically only the ``systematic''
DC-voltage $U_{\mathrm{DC}}$ 
contributes to the electrophoretic force
$\qq \vec E(\vec r,t)$,
while the effects of the 
``oscillating'' AC-driving $U_\mathrm{AC}\,\sin(\omega t)$
occur on time- and length-scales, which are of
no relevance for the effects we are interested in, and
practically average out to zero during one oscillation.
On the other hand, combining Eqs.~(\ref{1}), (\ref{6})
and (\ref{3})
we see that the time-averaged dielectrophoretic force
$\alpha \vec\nabla \vec E^2(\vec r,t)/2$
is proportional to
$\alpha \vec\nabla \vec E_1^2(\vec r) (U_\mathrm{DC}^2+U_\mathrm{AC}^2)/U_1^2
= \alpha \vec\nabla \vec E_1^2(\vec r) (1+U_\mathrm{DC}^2/U_\mathrm{AC}^2)(U_\mathrm{AC}^2/U_1^2)$.
For $|U_\mathrm{DC}| \ll |U_\mathrm{AC}|$
and sufficiently large
$U_\mathrm{AC}$,
the DC-contribution to the 
dielectrophoretic force
$\alpha \vec\nabla \vec E^2(\vec r,t)/2$
can always be safely neglected,
while the net effect of the AC-contribution over one oscillation
will still be significant.
In other words, electrophoresis is governed by the DC-
and dielectrophoresis by the AC-voltage.
In practice, these conditions are met by restricting
ourselves to relatively 
small DC-voltages (a few volts in the experiment),
while $U_\mathrm{AC}$ may be substantially larger
(several hundred volts). The AC frequencies $\omega/2\pi$
are restricted to
the range of several hundreds up to thousands of Hz
(a frequency of $\omega/2\pi=100\,\mbox{Hz}$
turned out to be optimal for performing the
experiments), thus also guaranteeing that the
quasi-static approximation for $\alpha$ is valid
(see also Sec.~2.1).

From Fig.~2 we can infer
that the electrophoretic 
forces alone would generate a net particle 
motion along the $x$-axis.
On the other hand,
the dielectrophoretic forces
give rise to ``particle traps'' (attractors,
deriving from local minima of the 
dielectrophoretic potential (\ref{3})).
Neglecting thermal noise effects for a moment,
we are thus faced with 
a competition between 
directed motion (governed by the DC-voltage)
and trapping (governed by the AC-voltage).
Depending on the relative magnitudes of
$U_\mathrm{DC}$ and $U_\mathrm{AC}$,
either the directed particle motion 
``wins'' or the particle stays trapped.
As a result, the entire parameter space spanned
by the DC- and the AC-voltage --
henceforth denoted as ``DC-AC-voltage plane'' --
can be divided into regions with
steady particle motion and regions 
where the particle stands still, see Fig.\ 3a.
Observing that the electrophoretic forces are 
proportional to $\qq U_\mathrm{DC}$ and the
dielectrophoretic forces proportional to 
$\alpha\,U^2_\mathrm{AC}$, the borderlines
between those regions are governed by
the equation
\begin{equation}
U^2_\mathrm{AC} = c\, |\qq U_\mathrm{DC}/\alpha|
\label{4}
\end{equation}
with a proportionality constant $c$ which depends on the
details of the microstructure
(``forming'' the spatial variations of the electric field),
the particle size, 
the sign of $\alpha$, and -- in case of asymmetric 
(ratchet shaped) obstacles --
on the direction of the electrophoretic forces
and thus on the sign of $\qq U_\mathrm{DC}$,
but otherwise is independent of
$\qq$, $\alpha$, $U_\mathrm{DC}$, and $U_\mathrm{AC}$.
\begin{figure}
\epsfxsize=\columnwidth
\epsfbox{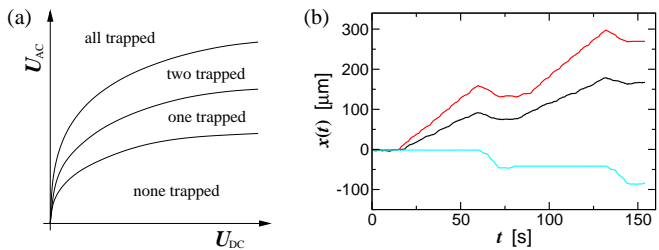}
\caption{Theoretical concept.
(a) Qualitative sketch of the
parameter space spanned by the DC- and AC-voltage
with $n$ moving and $N-n$ trapped 
particle species for $N=3$ (see main text).
The particle is trapped as long as
the combined deterministic electrokinetic effects result
in potential minima close to the obstacles
(see Fig.~2b and main text), i.e.\ at the border between two
different trapping regions (indicated by the solid
lines), the potential minima just disappear for
one of the particle species.
Thermal noise effects are neglected in that
picture; they would result in a \emph{finite}
trapping time which for that particle species
approaches zero close to border, but grows exponentially
with the depth of the potential minima when moving
away from the border into the trapping region towards
larger $U_\mathrm{AC}$ or smaller $U_\mathrm{DC}$.
For negative $U_\mathrm{AC}$,
the regions are symmetric (see (\ref{4}))
and hence omitted in the present plot.
For negative $U_\mathrm{DC}$,
Eq.\ (\ref{4}) implies an analogous symmetry
apart from a different proportionality constant $c$.
(b) Numerical solutions of (\ref{5})
for particles with diameters
1.1$\, \mu$m (black), 
1.9$\, \mu$m (red), and
2.9$\, \mu$m (blue),
moving in a periodically continued array 
of ratchet shaped obstacles as depicted in 
Fig. 2.
The external driving is specified
by Eqs.\ (\ref{1}), (\ref{6}) and by
Table 1, protocol a.
Remaining parameter values and simulation details:
see Appendices A.2-A.4 and A.5.
}
\label{fig3}
\end{figure}

The main consequence is that for two particles
species with different ratios $\qq/\alpha$ or 
different sizes,
the borderlines (\ref{4}) 
between motion and trapping are different.
(Here and in all what follows, we tacitly
focus on the case that $\qq$ and $\alpha$
have the same sign for all species).
As a consequence, the DC-AC-voltage plane
now divides into regions with (i) both particle
species moving, (ii) one moving and the other 
trapped, and (iii) both trapped.
Likewise, for $N>2$ particle species, the 
DC-AC-voltage plane exhibits regions with 
$n$ moving and $N-n$ trapped species for 
any $n=0,1,2,...,N$.
For an example with $N=3$, see Fig. 3a.

By exploiting the existence of these regions,
we will now construct a ``voltage protocol''
during which any particle of species $k$ 
is on average displaced by $\Delta x_k$ in 
$x$-direction, where $k=1,...,N$ labels
the species and the $\Delta x_k$ are arbitrary 
(positive or negative) multiples of 
the lattice constant in $x$-direction.
By repeating the corresponding protocol
periodically in time,
the average particle velocity (\ref{eq:v})
of species $k$ assumes the value
$\Delta x_k/\tau$, where $\tau$ denotes
the duration of one period of the voltage protocol.
By suitably choosing sign and modulus
of the $\Delta x_k$,
it is thus possible to make an arbitrarily 
selected subset of species move oppositely 
to all the others.
In other words, any preset 
separation task can be accomplished by 
properly tailoring the externally applied
DC- and AC-voltage.

Given the $\Delta x_k$ for $k=1,...,N$,
the announced protocol is composed of $N+1$
time intervals of lengths $\Delta t_n$, 
where $n=0,1,2,...,N$ and $\tau=\sum_{n=0}^N \Delta t_n$.
Throughout the $n$-th time interval, the
DC- and AC-voltages are kept at constant
values $U_\mathrm{DC}^{(n)}$ and 
$U_\mathrm{AC}^{(n)}$, respectively. 
As seen above, the actual values of
$U_\mathrm{DC}^{(n)}$ and $U_\mathrm{AC}^{(n)}$
can and will be chosen so that $n$ particle 
species are moving and $N-n$ are trapped.
Furthermore, the indices $k=1,...,N$
of the particle species can be chosen
without loss of generality so that
those with $k \leq n$ are moving 
during time interval $n$ and those
with $k>n$ stay trapped.
Put differently, particle $k$
(i.e. a particle of species $k$)
is moving during the time intervals 
with $n\geq k$ and is trapped when $n<k$.
Within the respective regions
of the DC-AC-voltage plane (see above),
$U_\mathrm{DC}^{(n)}$ 
and $U_\mathrm{AC}^{(n)}$
can still be freely chosen.
In particular,
$U_\mathrm{DC}^{(n)}$ may still be 
of either sign, and likewise the
transport direction of the 
moving species.

We first focus on particle $N$, which is 
only moving during the last time interval.
Hence, $\Delta t_N$, $U_\mathrm{DC}^{(N)}$, and 
$U_\mathrm{AC}^{(N)}$, can (and must) be 
chosen so that particle $N$ covers a distance 
$\Delta x_N$ during this time interval $N$.
During the same time interval,
all the remaining particles with 
$k=1,...,N-1$ are also moving.
Their respective displacements may be 
positive or negative and are denoted 
by $\delta x_k$.
Once we made our choice of $\Delta t_N$, $U_\mathrm{DC}^{(N)}$, 
and $U_\mathrm{AC}^{(N)}$ also the traveling 
distances $\delta x_k$ are uniquely fixed
and can be calculated quantitatively.

During all the remaining time intervals 
with $n=0,..,N-1$, particle $N$ stays 
trapped and thus its total displacement 
amounts to $\Delta x_N$, as desired.
On the other hand, the particles with 
$k=1,...,N-1$ now have to exhibit displacements
by $\Delta x_n'=\Delta x_n-\delta x_n$
during the remaining time intervals 
with $n=0,..,N-1$.
Along exactly the same line of reasoning, 
the existence of
appropriate values of $\Delta t_n$, $U_\mathrm{DC}^{(n)}$, 
and $U_\mathrm{AC}^{(n)}$ now follows by way
of induction for $n=N-1,N-2,...,1$.

The left-over time interval with $n=0$
may appear superfluous 
since all particles are trapped.
It is, however, essential for the separation
procedure to work efficiently.
In practice, there are various complications
and perturbations of the above idealized scheme,
most notably thermal noise and transient 
effects: after each ``jump'' of $U_\mathrm{DC}$ 
and $U_\mathrm{AC}$, every particle needs
a certain ``relaxation'' time to adapt 
its motion to the new circumstances 
(approach the new attractor of the dynamics), 
especially when switching from ``trapped'' 
to ``moving'' or vice versa.
All these effects would prevent a 
reliable time-periodic particle 
motion upon periodically repeating the 
protocol without the ``all-trapping'' 
time interval $n=0$.
Moreover, it guarantees that
the particles are indeed trapped
before species by species is released
from the traps during the next
cycle of the protocol.

We emphasize that the only indispensable
prerequisite of the above procedure is 
the existence of the different regions in 
the DC-AC-voltage plane (Fig.~3a), which in 
turn represents the generic case.
Different electrophoretic mobilities
of the particle species
or any other special system or particle 
properties are not required.
Accordingly, the parameters $\Delta t_n$, $U_\mathrm{DC}^{(n)}$, 
and $U_\mathrm{AC}^{(n)}$ are not yet uniquely
fixed and thus can still be
optimized with respect to the experimental
constraints and uncertainties.
Along these lines one typically ends up 
with alternating signs of $U_\mathrm{DC}^{(n)}$ 
and decreasing values of $|U_\mathrm{AC}^{(n)}|$
(the sign is irrelevant).
Moreover, in some cases one finds that less than $N+1$ 
time-intervals may already do the job, 
see Fig.~3b and Sec.~3.

In principle, by the above strategy one may
separate particle species of arbitrary 
sizes or $\qq/\alpha$ ratios in a 
rather broad dynamical range.
In particular, increasing noise effects for 
smaller particles can be kept under control
by applying larger voltages and hence
also enhancing the deterministic effects.
Likewise, reasonably small 
particle-to-particle variations 
within one species essentially lead to a 
shrinking of the pertinent regions in the 
DC-AC-voltage plane (cf. Fig.\ 3a)
and hence are still admissible.
However, too large variations lead to a disappearance 
of the regions and thus to a breakdown of our 
approach.
Analogous considerations apply to
deviations from a strictly periodic
microstructure.
On the other hand, a mixture of particles with
continuously distributed sizes or 
$\qq/\alpha$ ratios may still be sorted 
into certain subclasses, but a truly continuous 
separation is not possible.

Our general scheme would 
in principle also work for symmetric obstacles.
Ratchet shaped obstacles have the advantage 
of different $c$-values for positive
and negative DC-voltages in (\ref{4}), 
admitting much more efficient and robust 
separation in practice \cite{889}.
Likewise, the above mentioned
transients after each ``jump''  
of $U_\mathrm{DC}$ and $U_\mathrm{AC}$
can be exploited to further improve the efficiency 
of the procedure, especially
when taking into account thermal noise effects.
We finally note that somewhat similar concepts in the 
context of ratchet effects and particle sorting have 
been utilized previously in Refs.\ \cite{885,248,886}.

\section{Results}
To quantitatively verify these heuristic ideas, 
we perform detailed numerical simulations of the
model (\ref{5}), see Appendix A.5.
As a first example, we consider $N=3$ particle species.
Fig.~3a illustrates the corresponding regions in
the DC-AC-voltage plane with $n=0,1,2,3$ moving and 
$N-n$ trapped species.
Fig.~3b exemplifies the separation of
one species from all the others:
During an initial time-interval of $10\,$s
(see Table 1, protocol a), 
all particles are trapped. 
During the second interval of $50\,$s, two species are 
moving forward and one remains trapped.
During the last interval of $12\,$s, all move,
but now into the negative direction.
Upon periodically repeating the protocol
(two periods are shown), 
two species acquire a positive and one a 
negative average velocity.
(An interval with only one moving species is 
not needed for this specific separation task.)
By suitably tailored protocols (see Table 1), also
any other of the three particle species can be
made to migrate oppositely to the others,
as will be experimentally demonstrated 
below.
\begin{table*}
\small
\caption{Voltage protocols for the simulations in 
Figs.\ 3b, 6 and the experiments in Fig.\ 5.
Listed are the voltages $U_\mathrm{DC}$ and $U_\mathrm{AC}$ in volts
and the time-intervals $\Delta t_n$ in seconds
during which they were kept at those constant values.
The entire protocol $U(t)$ in (\ref{1})
is obtained by periodically
cycling through the listed sequence of time-intervals.
In all cases, the AC-driving frequency in (\ref{1}) 
is set to $\omega/2\pi=100\,\mbox{Hz}$.}
\begin{tabular*}{\textwidth}{@{\extracolsep{\fill}}ccc@{\hspace{7ex}}|ccc@{\hspace{7ex}}|ccc}
\hline
\multicolumn{3}{c|}{Protocol a} &
\multicolumn{3}{c|}{Protocol b} &
\multicolumn{3}{c} {Protocol c} \\
\hline
$\ \Delta t_n\ $ & $\ U_\mathrm{DC}\ $ & $\ U_\mathrm{AC}\ $ &
$\ \Delta t_n\ $ & $\ U_\mathrm{DC}\ $ & $\ U_\mathrm{AC}\ $ &
$\ \Delta t_n\ $ & $\ U_\mathrm{DC}\ $ & $\ U_\mathrm{AC}\ $ \\[1ex]
10 & 0 & 400 & 10 & 0 & 400 & 10 & 0 & 400 \\
50 & -15 & 320 & 90 & +15 & 400 & 40 & +15 & 360 \\
12 & +10 & 0 & 25 & -16 & 330 & 12 & -10 & 0 \\
\multicolumn{3}{c|}{} & 10 & +10 & 0 & \multicolumn{3}{c}{} \\
\hline
\end{tabular*}
\label{tab1}
\end{table*}

Before doing so, a second illustration with $N=5$ 
particle species is presented
with Fig.~4. Proceeding as detailed above, 
we have tailored voltage protocols for three
separation tasks which appeared to us particularly 
interesting and non-trivial:
Fig.~4a demonstrates the separation of the middle 
sized particles from the two bigger and the two 
smaller species.
Fig.~4b shows how the largest and smallest
particle species can be made to move oppositely 
to the three middle sized species.
The most challenging case is shown in Fig.~4c,
namely the disentanglement of the particles with 
1$\, \mu$m, 2$\, \mu$m, and 3$\, \mu$m diameter
from those with 1.5$\, \mu$m and 2.5$\, \mu$m diameter.
\begin{figure}
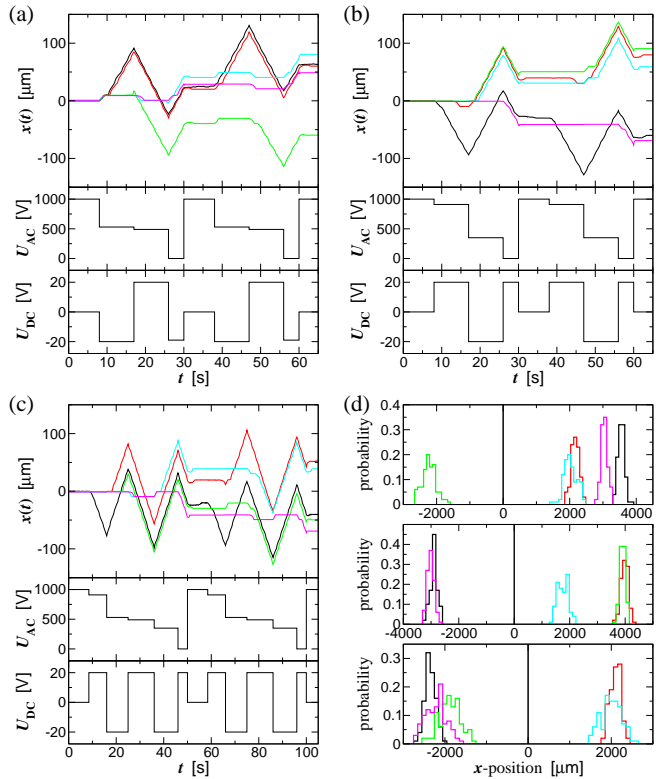

\begin{tabular}{c@{\hspace*{0.3cm}}c}
\epsfxsize=0.48\columnwidth
\epsfbox{fig4a.eps}
&
\epsfxsize=0.48\columnwidth
\epsfbox{fig4b.eps}
\\ 
\epsfxsize=0.48\columnwidth
\epsfbox{fig4c.eps}
& 
\epsfxsize=0.48\columnwidth
\epsfbox{fig4d.eps}
\end{tabular}
\caption{Numerical solutions for 5 particles with diameters
1$\, \mu$m (black), 
1.5$\, \mu$m (red),
2$\, \mu$m (green),
2.5$\, \mu$m (blue),
and 3$\, \mu$m (magenta).
(a): Separation of the 2$\, \mu$m particles 
(middle size) from all the other.
(b): Separation of the smallest and largest 
(1$\, \mu$m and 3$\, \mu$m) particles from all the 
others (3 intermediate diameters).
(c): Separation of the particles with 
1$\, \mu$m, 2$\, \mu$m, and 3$\, \mu$m diameter
from those with 1.5$\, \mu$m and 2.5$\, \mu$m diameter.
The lower panels specify the voltage protocols
used to achieve the separation tasks shown in the 
respective upper panels.
In each case, two periods of the voltage protocols are shown.
The explicit algorithm to devise those protocols
is given in the main text.
Further details and parameter values are as in 
Fig.~3b, see also Appendices A.2-A.5.
(d): Spatial distributions of 100 particles of each species obtained
after simulating 100 periods of the voltage protocols. The upper panel
corresponds to the separation task from (a), the middle panel to (b),
and the lower panel to (c).
}
\label{fig4}
\end{figure}

Closer inspection of Figs.~4a-c reveals that the particle
motion does not always precisely follow the displacement pattern
anticipated in the construction of the voltage
protocol as described above (Sec.~2.3),
but rather shows significant variations between different periods of the protocol.
For instance, the $2\,\mu\mbox{m}$ (green) particle in Fig.~4c moves in four (out
of the six) voltage steps during the first period and only in three
during the second period.
Likewise, the $2.5\,\mu\mbox{m}$ (blue) particle shows a net
displacement of about $30\,\mu\mbox{m}$ during the first cycle but
performs no net movement in the second cycle.
These variations are due to thermal noise effects,
which have a strong impact
on the escape process out of the dielectrophoretic traps,
because these are located close to stagnation points of the
DC electric field (see Fig.~2).
Due to this variability of the net
displacement per voltage cycle, particles of the same 
species are spread out along their direction of systematic 
motion farther than by pure thermal diffusion. 
Such enhanced dispersion notwithstanding, efficient 
particle fractionation based on transport in opposite
directions is demonstrated by Fig.~4d
for all three separation tasks from Figs.~4a-c.
Note that even within the subsets of particle species,
that have been selected to migrate in one specific
direction, different migration velocities
for different species and thus (partial) separation
may be observed (e.g.\ in the first panel of Fig.~4d).
This ``side-effect'' results
from performing the separation by less than
$N+1$ steps in the voltage protocol, so that
the displacements
$\Delta x_k$ have not been determined and optimized independently
for all the species in the subset 
(see the algorithm for constructing the voltage protocols,
as described in Sec.~2.3), and from the fact
that the deviations from the anticipated
displacement patterns depend on particle properties and are
thus different for different particle species within one subset.

Next, we turn to the experimental test of
our theoretical concepts by means of the setup as
detailed above in Fig.\ 1 and in Appendix A.1.
Fig.\ 5 summarizes our main findings, demonstrating that 
for a colloidal suspension containing three different species
of polystyrene beads it is indeed possible to control the
transport directions to the extent that any one of 
the three species can be made to propagate oppositely 
to the other two species.
\begin{figure}[b]
\epsfxsize=\columnwidth
\epsfbox{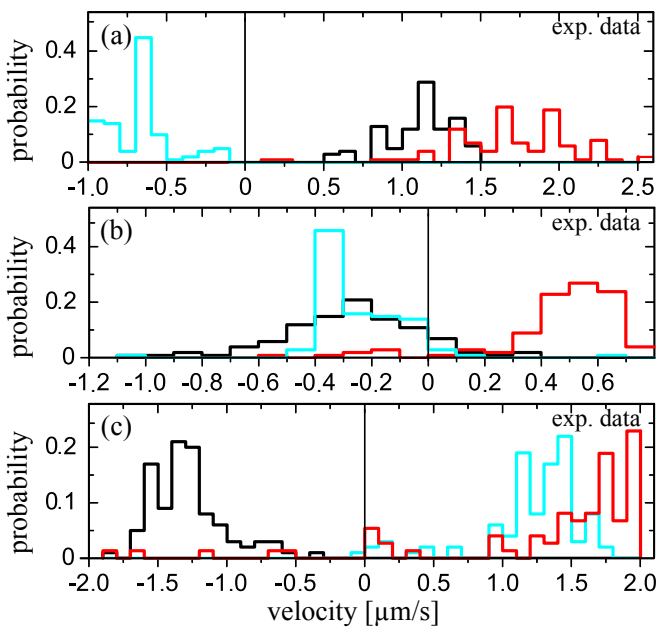} 
\caption{Experimentally measured velocity histograms for 
$N=3$ particle species
(polystyrene beads with diameters
1.1$\, \mu$m (black), 
1.9$\, \mu$m (red), and
2.9$\, \mu$m (blue)),
simultaneously moving in the structured 
microfluidic ratchet device from Fig.\ 1.
(a)-(c) show the results for the three 
different time-periodic voltage protocols
from Table 1.
For each histogram, the displacement of 100 beads
in $x$-direction during one time-period $\tau$ of 
the driving voltage has been measured and
interpreted as their net velocity during that
driving period, i.e.\ an average over the histogram
for a specific particle species gives the
average velocity (\ref{eq:v}) for that species.
In (a), the 2.9$\,\mu$m beads move oppositely 
to the other two species with very high
reliability, and analogously for the 
1.9$\,\mu$m beads in (b) and the
1.1$\,\mu$m beads in (c).
}
\label{fig5}
\end{figure}

Finally, Fig.~6 shows numerical simulations,
obtained along the very same lines as in Fig.~3b, except that
the so generated numerical data have now been further 
processed exactly as the experimental data in Fig.~5a.
These numerical results in Fig.~6 compare very 
well with the experimental 
counterparts from Fig.\ 5a.
The fact that the velocity spreads in Fig.~5a are not
significantly larger than in Fig.~6 implies that these
experimental spreads are mainly caused by the thermal
noise effects discussed above and not by variations of 
$\qq$, $\alpha$, and $\friction$ for beads of the same species.
The minor remaining differences between 
experiment and simulations are probably due to nonlinear (induced-charge)
electrokinetic effects \cite{887,736},
which we have neglected in our model (\ref{5}).
Furthermore, it can not be fully excluded that
(small) Joule heating effects \cite{xuan08}
affect the particle motion,
although we do not observe any of the typical
indicators for Joule heating during the experiments,
like vortex-like
flow patterns at the constrictions created by the obstacles
\cite{hawkins10,sridharan11}
or the generation of gas bubbles. 
\begin{figure}
\epsfxsize=\columnwidth
\epsfbox{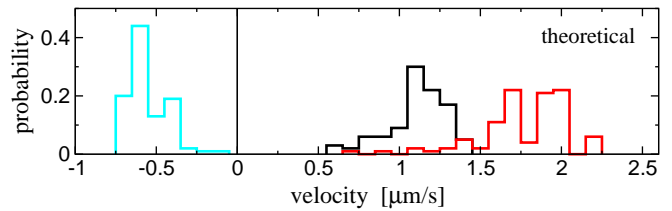} 
\caption{Velocity histograms by numerically simulating 
the experiment from Fig.\ 5a according to 
(\ref{1}), (\ref{5}), and Table 1.}
\label{fig6}
\end{figure}

\section{Conclusions}
In summary, we have put forward a new concept 
for particle sorting in a microstructured ratchet device 
by exploiting a subtle competition
between electrophoretic and dielectrophoretic forces.
Within one and the same device, by tailoring the applied
voltage protocol, one or several arbitrarily selected particle 
species can be made to propagate oppositely to all
the other particle species.
The choice of those particular species can be easily 
changed and adapted at any time, even while the 
experiment is running, by simply adapting 
the applied voltage protocol.

Such versatile particle transport is achieved by
exploiting all ``elementary ways of symmetry breaking''
in the context of ratchet effects \cite{871,361,125,872}, namely
spatial symmetry breaking,
temporal (or dynamical) symmetry breaking,
and application of a bias force.
Any demand to maintain certain
symmetries would amount to an unnecessary constraint,
whereas their combined breaking allows for the 
required flexibility and efficiency \cite{889}.

Our new separation concept has been devised theoretically and 
realized experimentally within a particle density regime, 
in which particle interactions other than excluded volume 
effects are negligible
and in which the number of dielectrophoretic particle traps is
larger or comparable to the total number of particles in the microfluidic device.

In the future, experiments with
biological complexes ranging
from single biopolymers, like DNA or proteins, to whole cells,
as well as non-biological objects, like carbon-nanotubes,
can be envisioned.
In spite of the fact that dielectrophoretic forces are
theoretically expected to strongly decrease with particle size
(cf. Eq.\ (\ref{8})), the feasibility of manipulating 
single proteins by electrodeless dielectrophoresis has 
already been experimentally demonstrated e.g. in
\cite{cla05,cla07,lap08}.
Moreover, separation may be performed not only according to 
size but also according to molecule conformation or particle 
shape \cite{reg10}
and thus represents a considerable advance beyond 
conventional separation techniques, 
like (pulsed-field) gel electrophoresis.
A further interesting direction for future research
are hydrodynamic and other interaction effects as well as
possible modifications of the separation mechanism, arising
for substantially higher particle densities than in our
present experiments.

This work was supported by Deutsche Forschungsgemeinschaft 
under SFB 613 and RE1344/5-1

\appendix
\section{Materials and Methods}
\subsection{Experimental details}
The entire microdevice (see third paragraph and Fig.~1)
was cast with poly(dimethylsiloxane) (PDMS),
mounted on a PDMS coated coverslip after a treatment
with an oxygen plasma for 30s,
and coated with a commercial triblock 
copolymer (F108, 500 $\mu$M) for 2 hours \cite{107}.
Spherical polystyrene beads (Molecular Probes) with diameters
1.1$\,\mu$m, 1.9$\,\mu$m, and 2.9$\,\mu$m 
at a concentration of 50 - 500 particles per species
and nanoliter were immersed in a 10 mM phosphate buffer (pH 8.3)
containing 200$\mu$M Tween 20.
Hence, observable hydrodynamic or other 
interaction effects were always negligibly rare.
For a more detailed description of our PDMS chip fabrication,
particle and microstructure pre-treatment,
we refer to \cite{134,850}.
Note, that the F108 coating of the
microstructure strongly suppresses electroosmotic
effects \cite{107}, as well as adhesion of the
polystyrene beads to the PDMS surfaces
(in the experiments less than 1\% of the particles
got stuck to the PDMS structure).
Particle tracking by video microscopy was carried
out with a standard fluorescence microscope.
We recorded the motion of the
particles within a field of view of about $825\,\mu\mbox{m} \times 310\,\mu\mbox{m}$
with a frame-rate of 5$\,\mbox{fps}$.
Using the software ImageJ 1.43 (with the plugin MtrackJ) \cite{abr04,mei},
particle trajectories were extracted from the
recorded videos with a spatial resolution of $0.6\,\mu\mbox{m}$
(both in $x$ and $y$ direction) and a temporal precision
of $0.2\,\mbox{s}$.
The three different particle species were easily 
distinguishable by their size and/or fluorescent 
labeling (Fig.~1c). The histogram plots in Fig.~5
were obtained from the trajectories
by determining the particle
displacements
during one period of the voltage
protocol; the so obtained 
velocities have a relative error below 10\%.

\subsection{Friction coefficient}
For a spherical particle with diameter $d$
in a fluid of viscosity $\viscosity$ and infinitely
far boundaries,
the friction coefficient $\friction$ in Eq.\ (\ref{5}) could 
be approximated by Stokes' formula $\friction_{Stokes}=6\pi\viscosity (d/2)$.
Here, we go one step beyond this approximation, accounting
very roughly for the hydrodynamic interactions with 
the microstructure by experimentally determining the
diffusion coefficient $D$ and then 
inferring $\friction$ from Einstein's relation $D=kT/\friction$.
Those experimentally estimated values were
$D=0.202\pm 0.002\,\mu$m$^2$/s for the 1.1$\,\mu$m particles,
$D=0.131\pm 0.005\,\mu$m$^2$/s for the 1.9$\,\mu$m particles, and
$D=0.082\pm 0.007\,\mu$m$^2$/s for the 2.9$\,\mu$m particles.
The resulting $\friction$-values are close to $1.7\cdot\friction_{Stokes}$
in all three cases. For the five particle species simulated 
in Fig.\ 4 we thus assume along the same line of reasoning that
$\friction=1.7\cdot\friction_{Stokes}$.

\subsection{Effective charge}
The effective charge $\qq$ from 
(\ref{2}) depends in 
a very complicated way on the particle material, 
the buffer solution, and the surface coating 
of particle and microstructure \cite{squ05,834,436}.
In principle, one would expect that particles 
which only differ by their diameter have the same
zeta potential, and hence the effective charge
(\ref{2}) would inherit the linear dependence
on the particle diameter from the friction 
coefficient $\friction$.
In practice, even nominally identical particles
but purchased at different times or from different 
companies are found to exhibit different electrophoretic 
mobilities due to somewhat differing production 
conditions and also due to ageing effects
of the surface coating
(induced by hardly controllable
(bio-)chemical reactions at the 
particle-fluid interface).
In view of these practical problems, we have directly
estimated $\qq$ by fitting
the model (\ref{5}) to the experimentally 
observed particle motion, resulting in
$\qq= - 0.055\cdot 10^{-15}\,\mbox{C}$ for the 1.1$\,\mu$m particles,
$\qq= - 0.17\cdot 10^{-15}\,\mbox{C}$ for the 1.9$\,\mu$m particles, and
$\qq= - 0.50\cdot 10^{-15}\,\mbox{C}$ for the 2.9$\,\mu$m particles.

For the five particle species simulated in Fig.\ 4,
we assumed -- in the absence of experimental data --
the above mentioned idealized theoretical situation 
of identical zeta potentials and thus identical 
Helmholtz-Smoluchowski velocities (\ref{7}).
In the concomitant linear dependence of
$\qq$ on the particle diameter,
the proportionality constant was chosen so 
as to yield a net velocity of 6$\,\mu$m/s in 
$x$-direction when driven by a purely static 
voltage of  $U_{\mathrm{DC}}=10\,$V.
This velocity is close to those we observed
for the three experimental particle species.

We remark that the electroosmotic fluid flow
$\vec v(\vec r)$ of the ionic liquid exhibits
a remarkable ``similitude'' to the
electrophoretic Helmholtz-Smoluchowski relation (\ref{7}),
namely $\vec v(\vec r)=-\vec E(\vec r) \epsilon\zeta'/\viscosity$,
where $\zeta'$ is the zeta potential of the 
microstructure, see \cite{436,san01} and further
references therein.
The effect of this electroosmotic flow on the particle 
can be readily incorporated into the above mentioned
effective electrostatic force $\qq \vec E(\vec r)$ 
by replacing $\zeta$ in the definition (\ref{2})
by $\zeta-\zeta'$.
In our actual experiment, those electroosmotic 
effects are, however, strongly suppressed by
the F108-coating of the microstructure (see above).

\subsection{Effective polarizability}
While the particle polarizability is in 
general frequency dependent, experimentally
we always work in the quasi-static regime 
($\omega/2\pi=100\,$ Hz), and hence
our $\alpha$ always refers to the zero frequency limit. 
Invoking the basic formula (\ref{8})
and approximating the medium's permittivity $\epsilon$ by 
that of water, $\epsilon = 80\epsilon_0$ 
($\epsilon_0$ being the vacuum dielectric constant),
one finds that
$\alpha = -7.4\cdot 10^{-28}\,$ Fm$^2$ for the 1.1$\,\mu$m particles,
$\alpha = -38 \cdot 10^{-28}\,$ Fm$^2$ for the 1.9$\,\mu$m particles, 
and $\alpha = -136\cdot 10^{-28}\,$ Fm$^2$ for the 2.9$\,\mu$m 
particles.
The negative sign of $\alpha$ is in agreement with
previous experiments under comparable conditions 
and has been tacitly anticipated in Fig.~2b.

However, the basic formula (\ref{8}) neglects, among others, 
the polarizability of the counter-ion cloud (Debye layer)
around the dielectric sphere, which, in turn, is much 
more difficult to estimate theoretically.
Like for the above discussed effective charges,
we therefore estimated the relevant effective polarizabilities
by directly fitting the experimentally observed particle motion
to the theoretical dynamics from Eq.\ (\ref{5}),
resulting in 
$\alpha = -4.0\cdot 10^{-28}\,$ Fm$^2$ for the 1.1$\,\mu$m particles,
$\alpha = -27 \cdot 10^{-28}\,$ Fm$^2$ for the 1.9$\,\mu$m particles, and
$\alpha = -126\cdot 10^{-28}\,$ Fm$^2$ for the 2.9$\,\mu$m particles.
The slight systematic deviations from (\ref{8}) are remarkably small.
For this reason and due to the lack of experimental data,
the polarizabilities for the five particle species simulated 
in Fig.\ 4 were chosen according to formula (\ref{8}) without
any further corrections.

\subsection{Simulation details}
For numerical simulations, the model (\ref{5})
was discretized according to the Euler algorithm
with a time step of $dt=10\,\mu\mbox{s}$.
The electric field $\vec E_1(\vec r)$
and the dielectrophoretic ``field'' $\vec\nabla\vec E_1^2(\vec r)$
was computed on a spatial grid with step sizes $0.02\,\mu\mbox{m}$
within one basic ``unit cell'' of the
microstructure around an obstacle (cf.\ Figs.~2a,b)
by using the COMSOL Multiphysics software
suite. This was done once before performing the dynamic
simulations based on (\ref{5}). The actual electrophoretic and
dielectrophoretic forces at the current particle
position during the simulations were determined 
by periodically continuing the fields within this ``unit cell''
(in $x$ and $y$ direction),
by interpolating linearly between the relevant grid points,
and by taking into account (\ref{6}).
The term $\vec F(\vec r)$ in (\ref{5}), representing idealized hard-wall
interactions with the microstructure, was treated
according to the so-called rejection scheme \cite{behringer11}.
The triangular posts of the microstructure were
approximated by the shape shown in Fig.~2.

To obtain the simulation results of Fig.~4, 100 independent realizations
were computed for each particle species over 100 periods
of the respective voltage protocol (the Figs.~4a-c show one typical realization
for each particle species during the first two driving periods).
In Fig.~6, the results of 100 independent realizations
for each particle species are shown, evaluated over one period of
the voltage protocol (in accordance with the
processing of the experimental data).

\footnotesize{

}





\end{document}